\documentclass{pasj00}

\SetRunningHead{Q. Wada et al.}{Suzaku Detection of an X-ray Pulse from the B\textit{e}/NS System in the SMC}

\Received{}
\Accepted{}
\Published{}

\begin{document}
\title{Detection of a 522~s Pulsation from the Transient X-ray Source\\
Suzaku\,J0102.8--7204 (SXP\,523) in the Small Magellanic Cloud}
\author{
Qazuya~\textsc{Wada},\altaffilmark{1,2}
Masahiro~\textsc{Tsujimoto},\altaffilmark{1}
Ken~\textsc{Ebisawa},\altaffilmark{1,2}
Eric D.~\textsc{Miller}\altaffilmark{3}
}
\altaffiltext{1}{
Japan Aerospace Exploration Agency, Institute of Space and Astronautical Science,\\
3-1-1 Yoshino-dai, Chuo-ku, Sagamihara, Kanagawa 252-5210
}
\email{wada@astro.isas.jaxa.jp}
\altaffiltext{2}{
Department of Astronomy, Graduate School of Science,\\
The University of Tokyo, 7-3-1 Hongo, Bunkyo-ku, Tokyo 113-0033
}
\altaffiltext{3}{
Massachusetts Institute of Technology, Kavli Institute for Astrophysics and Space
Research, \\
77 Massachusetts Ave 37-582G, Cambridge, MA 02139, USA
}

\maketitle

\begin{abstract}
 During a routine calibration observation of 1E\,0102.2--7219 in the Small Magellanic
 Cloud (SMC) carried out in October 2012 for the Suzaku satellite, we detected a
 transient X-ray source at (RA, Dec) $=$ (\timeform{01h02m47s}, \timeform{-72d04m54s})
 in the equinox J2000.0 with a positional uncertainty of $\sim$1\farcs4. We conducted
 a temporal and spectral analysis of the source and found a coherent pulse signal with
 a period of 522.3 $\pm$ 0.1~s, and a featureless spectrum described by a single
 power-law model with a photon index of $1.0^{+0.1}_{-0.1}$ and a 0.5--10~keV luminosity
 of $\sim$8.8$\times$ 10$^{35}$ erg~s$^{-1}$ at an assumed distance of 60~kpc. The Suzaku
 source is likely to be the counterpart of 2XMM\,J010247.4--720449, which has been
 observed several times, including during outburst by Swift. Based on the X-ray
 characteristics in our data, as well as the transient record and optical and
 near-infrared features in the literature, we conclude that this source is a high-mass
 X-ray binary pulsar with a B\textit{e} star companion in the SMC, which is known to harbor
 an exceptionally large ($\sim$80) number of such sources in comparison to our Galaxy.
\end{abstract}

\section{Introduction and Summary}
The Small Magellanic Cloud (SMC) has a mass of only $\sim$10$^{-2}$ of our own Galaxy,
yet it hosts a comparable number of high-mass X-ray binary pulsars \citep{sturm11}.  All
but one of them are binary systems of a neutron star (NS) and a B\textit{e} star
\citep{coe12}. The extreme over-density of such sources infers an elevated star-forming
activity about 25--60~Myr ago \citep{yokogawa03,antoniou10}.

Almost all these sources are recognized when they go outburst around the periastron
passages in a highly eccentric orbit when the mass accretion rate to the NS
increases. During the outburst, an X-ray pulsation of 1--1000~s \citep{knigge11} is
often detected, which gives a conclusive piece of evidence for identifying such
sources. The orbital cycle is long (10--1000 days; \cite{knigge11}), suggesting that
there are an even larger number of B\textit{e}/NS binary systems in the SMC.

In order to search for such sources yet to be discovered, X-ray survey observations were
conducted several times using the ROSAT \citep{haberl00,sasaki00}, ASCA
\citep{yokogawa03}, XMM-Newton \citep{haberl12}, and Chandra \citep{mcgowan08} X-ray
observatories. X-ray monitoring observations were also made with the RXTE
\citep{galache08}. All these observations contributed to the detection of pulse signals
from transient sources and to the localization for follow-up optical spectroscopy to
reveal their B\textit{e}/NS nature.

Less explored than survey and monitoring observations are routine calibration
observations in the SMC. The galaxy hosts 1E\,0102.2--7219 (E0102 hereafter), a young
super-nova remnant, which is among the most often observed sources for calibrating
in-orbit X-ray instruments because of its soft and line-dominated spectrum, stable flux,
and good visibility throughout the year. The X-ray Imaging Spectrometer (XIS;
\cite{koyama07}) onboard the Suzaku satellite \citep{mitsuda07} has observed this source
once every few months for the purpose of monitoring the energy gain and the
contamination build-up. As of writing, the XIS observed E0102 55 times with a total
integration time of 1.7~Ms since 2005 August. In terms of the area
(18\arcmin$\times$18\arcmin\ for the XIS) times the exposure (20--30~ks every time), the
depth of coverage is comparable to the SMC survey performed with the European Photon
Imaging Counter (EPIC) on XMM-Newton \citep{haberl12}. Therefore, such data sets provide
a unique opportunity for searching bright transient sources and revealing their
long-term behavior, as is illustrated in \citet{haberl05,eger08,takei08}.

\medskip

In this Letter, we present the detection of a transient source during two of our E0102
calibration observations with the XIS. We detected a coherent pulse of 522~s and
obtained a power-law spectrum \citep{wada12}. The pulse period was also confirmed by a
successive XMM-Newton observation \citep{sturm13}. Together with the drastic X-ray flux
change and the features of the counterpart in the longer wavelengths found in previous
studies, we conclude that this source is another B\textit{e}/NS binary in the SMC.

\section{Observations \& Data Reduction}
We conducted a calibration observation of E0102 on 2012 October 29, in which we
recognized a bright transient source. The XIS is equipped with four X-ray CCD cameras at
the focii of four X-Ray Telescope (XRT; \cite{serlemitsos07}) modules, and has an
imaging-spectroscopic capability in the 0.2--12.0~keV band. The four sets of cameras and
telescopes are co-aligned with each other to provide a 18\arcmin$\times$18\arcmin\ field
of view with a telescope half power diameter (HPD) of $\sim$2\arcmin, independent of
X-ray energies. Three of the CCD cameras (XIS0, 2, and 3) carry front-illuminated (FI)
devices, while the remaining one (XIS1) carries a back-illuminated (BI) device. The
devices have an energy resolution of 180~eV as a full width at half maximum (FWHM) at
5.9~keV as of the observation date. The FI and BI devices are superior to each other in
the hard and the soft band response, respectively. The entire XIS2 and a part of the
XIS0 cameras are dysfunctional since 2006 November and 2010 December, respectively, due
to putative micro-meteorite hits, thus we used the remaining parts of the cameras. The
total effective area amounts to $\sim$800~cm$^{2}$ at 1.5~keV at the field center.

We operated the XIS in the normal clocking mode with a frame time of 8~s. Events were
removed when they were taken during South Atlantic Anomaly (SAA) passages, elevation
angles from the day Earth by $\le$5\degree\ and from the night Earth by
$\le$20\degree. The net exposure time was 32.1~ks. We used the HEADAS software
package\footnote{See http://heasarc.gsfc.nasa.gov/docs/software/lheasoft/ for details.}
version 6.12 for the X-ray data reduction throughout this paper.

\section{Analysis}
\subsection{Image Analysis}
\begin{figure}[ht]
 \begin{center}
  \FigureFile(75mm,75mm){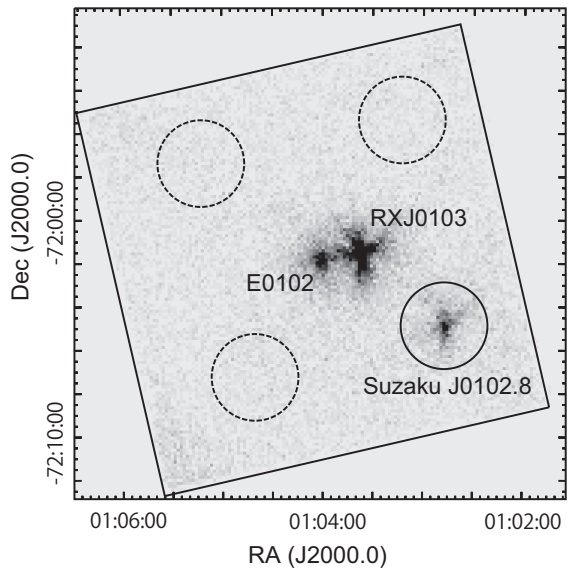}
 \end{center}
 \caption{XIS image in the 2.0--10~keV band after the astrometric correction. Events
 taken with XIS0, 1, and 3 are merged. The source and background extraction regions are
 shown with solid and dashed circles.}
 \label{f1}
\end{figure}

Figure~\ref{f1} shows the XIS image in the 2.0--10~keV band. E0102 was observed at the
center of the field. Another source RX\,J0103.6--7201 (RXJ\,0103) can be found 1\farcm9
to the east. E0102 is a supernova remnant, which is predominantly bright below 2~keV,
while RXJ\,0103 is a high-mass X-ray binary, which is comparable in
the hard-band brightness with E0102. Yet another source is apparent at a 6\farcm4
distance from E0102 displaced to the south west. This is a transient source as we did
not recognize such a source in the previous XIS observations of E0102.

E0102 is an extended source, while RXJ\,0103 is a point-like source, so we used the
latter for the astrometric correction by fitting the intensity profile with the point
spread function. After shifting by ($\Delta$RA, $\Delta$Dec) $=$ (27\farcs7, 2\farcs0),
we derived the position of the transient source to be (RA, Dec) $=$
(\timeform{01h02m47s}, \timeform{-72d04m54s}) in the equinox J2000.0 with an uncertainty
of $\sim$1\farcs4. We named the source Suzaku\,J0102.8-7204.
We extracted source events from a 2\arcmin\ radius circle around the source, and the
background events from three circles of the same radius at an equal distance from
E0102 (figure~\ref{f1}).

\subsection{Temporal Analysis}
\begin{figure}[ht]
 \begin{center}
  \FigureFile(80mm,60mm){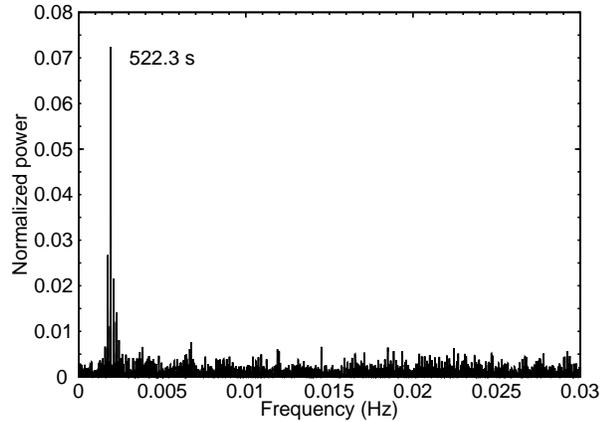}
 \end{center}
 \caption{Power spectral density. A peak is found at 1.91 $\times$ 10$^{-3}$ Hz.}
 \label{f2}
\end{figure}

\begin{figure}[ht]
 \begin{center}
  \FigureFile(80mm,60mm){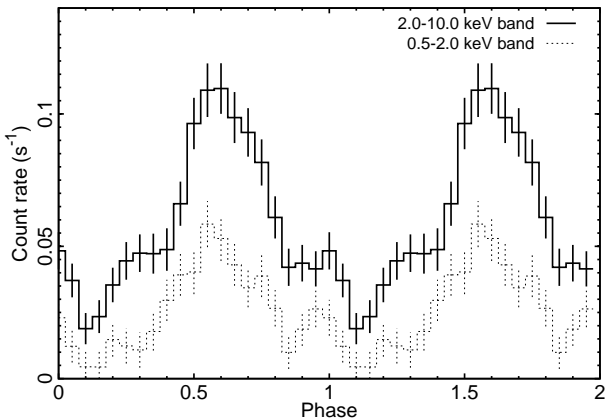}
 \end{center}
 \caption{Folded and background-subtracted light curve for two cycles in the 0.5--2
 (dashed) and 2--10 (solid) keV bands.}
 \label{f3}
\end{figure}

We first constructed the light curve of the transient source and confirmed that there
were no flaring events nor any trends of flux variation during the observation. After a
barycentric correction of the photon arrival times, we conducted a period search using
the generalized Lomb-Scargle method \citep{zechmeister09}. Figure~\ref{f2} shows the
power spectral density using the events in the 0.5--10.0~keV band in the 0.01--31.25~mHz
range. We found a peak at 1.91 $\times$ 10$^{-3}$ Hz with a statistical probability of
$<$10$^{-10}$ for the peak to be a background fluctuation. We determined the fundamental
period of 522.3 $\pm$ 0.1~s. Following the naming convention by \citet{coe05}, the
source is alternatively called SXP\,523\footnote{It should be SXP\,522, but we stick to
the original naming by \citet{haberl12b} to avoid confusion.}. Figure~\ref{f3} shows the
background-subtracted light curve folded by the period separately for the soft
(0.5--2~keV) and hard (2--10~keV) bands. No significant change in the period was found
between the two halves of the observation.

\subsection{Spectral Analysis}
\begin{figure}[ht]
 \begin{center}
  \FigureFile(80mm,60mm){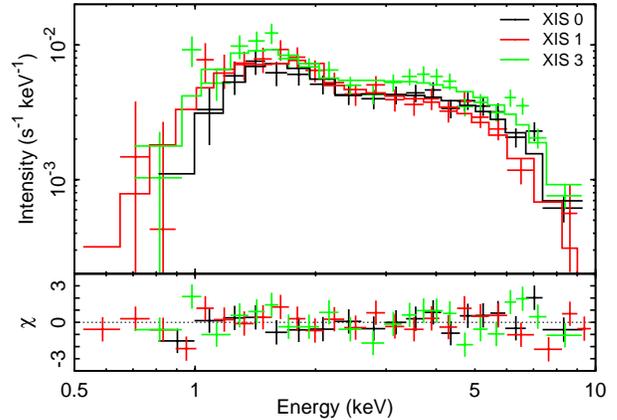}
 \end{center}
 \caption{XIS0 (black), 1 (red), and 3 (green) spectra in the 0.5--10.0~keV band. The
 upper panel shows the data with the crosses and the best-fit power-law model with the
 solid lines. The lower panel shows data residuals from the best-fit model.}
  \label{f4}
\end{figure}

Figure~\ref{f4} shows the background-subtracted X-ray spectrum. In order to fit the
spectrum, we generated the detector and telescope response files using the
\texttt{xisrmfgen} and \texttt{xissimarfgen} \citep{ishisaki07} tools, respectively.

The spectrum is featureless, thus we fitted it with a power-law model
attenuated by a photoelectric absorption model (\texttt{tbabs}:
\cite{wilms00}) using the Xspec software version 12.7.1. The free parameters are the
power-law slope ($\Gamma$), flux in the 0.5--10~keV band ($F_{\mathrm{X}}$), and the
absorption column in the SMC ($N_{\mathrm{H}}^{\mathrm{SMC}}$). The Galactic absorption
column ($N_{\mathrm{H}}^{\mathrm{Gal.}}$) was fixed to 5.5$\times$10$^{20}$~cm$^{-2}$
\citep{dickey90}. The metal abundance in the SMC was set to 0.2 solar 
\citep{russel92}. The spectrum was fitted quite well with such a simple model. The
best-fit parameters are shown in table~\ref{t1}. The absorption-corrected luminosity in
the 0.5--10~keV band is $\sim$8.8$\times$ 10$^{35}$ erg~s$^{-1}$ assuming a distance of
60~kpc to the SMC \citep{mathewson85}.

\begin{table}[ht]
 \begin{center}
  \caption{Best-fit parameters for the X-ray spectra.\footnotemark[$*$]}
  \label{t1}
  \begin{tabular}{lcc}
   \hline
   \hline
   Par. (unit) & 2012 Oct 29 & 2012 June 25 \\
   \hline
   $N_{\mathrm{H}}^{\mathrm{Gal.}}$\footnotemark[$\dagger$] (10$^{20}$ cm$^{-2}$) & 5.5 (fixed) & 5.5 (fixed) \\
   $N_{\mathrm{H}}^{\mathrm{SMC}}$\footnotemark[$\dagger$] (10$^{22}$ cm$^{-2}$)  & $2.3^{+0.7}_{-0.6}$ & $0.5^{+1.3}_{-0.5}$ \\
   $\Gamma$ & $1.0^{+0.1}_{-0.1}$ & $1.3^{+0.2}_{-0.2}$\\
   $F_{\mathrm{X}}$\footnotemark[$\ddagger$] (10$^{-12}$ erg~s$^{-1}$~cm$^{-2}$) &
       $2.0^{+0.1}_{-0.1}$ & $0.5^{+0.1}_{-0.1}$\\
   $\chi^{2}_{\rm{red}}$/dof\footnotemark[$\S$] & 1.01/59 & 1.30/29 \\
   \hline
   \multicolumn{3}{@{}l@{}}{\hbox to 0pt{\parbox{85mm}{
   \footnotesize
   \par \noindent
   \footnotemark[$*$] The errors indicate a 1$\sigma$ statistical uncertainty.\\
   \footnotemark[$\dagger$] $N_{\mathrm{H}}^{\mathrm{Gal.}}$ and $N_{\mathrm{H}}^{\mathrm{SMC}}$ are the hydrogen-equivalent column density of the Galactic foreground and the SMC, respectively.\\
   \footnotemark[$\ddagger$] The 0.5--10.0~keV band flux not corrected for the absorption.\\
   \footnotemark[$\S$] The reduced $\chi^{2}$ ($\chi^{2}_{\rm{red}}$) and the
   degrees of freedom (dof). \\
   }\hss}} 
  \end{tabular}
 \end{center}
\end{table}

\subsection{Other Data}
We retrieved all the previous E0102 observations with the XIS and actually found a
significant detection on 2012 June 25 with a count rate $\sim$1/4 of that recorded in
2012 October. We derived the pulse period of 521.8 $\pm$ 0.4~s by epoch folding and the
spectral parameters in table~\ref{t1}.

\section{Discussion \& Conclusion}
\begin{figure}[ht]
 \begin{center}
  \FigureFile(80mm,80mm){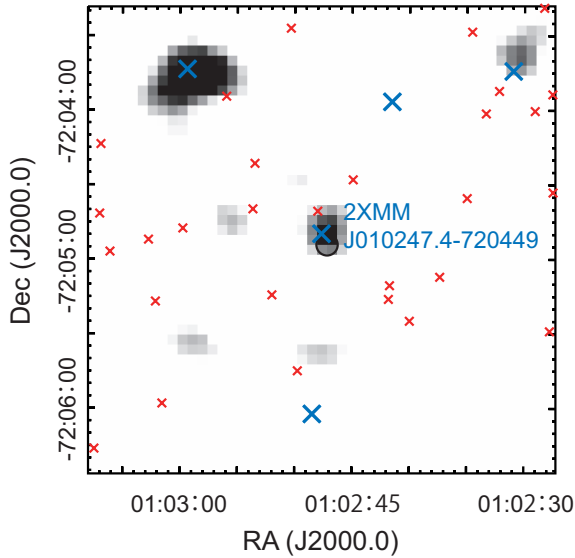}
 \end{center}
 \caption{XMM-Newton EPIC image in the 0.5--10~keV band taken on 2000 April 16. The
 positions of XMM-Newton catalog \citep{watson09} and 2MASS sources are shown
 respectively with blue and red crosses. The 3$\sigma$ positional error for the Suzaku
 source is shown with the circle at the center.}
 \label{f5}
\end{figure}

The region around E0102 has been observed numerous times also by other X-ray
instruments. Figure~\ref{f5} shows an XMM-Newton image around Suzaku\,J0102.8--7204. An
XMM-Newton source named 2XMM\,J010247.4--720449 is found within the Suzaku error circle.
2XMM\,J010247.4--720449 is a transient source and very likely to be the same with the
Suzaku source \citep{haberl12b}.

\begin{figure}[ht]
 \begin{center}
  \FigureFile(80mm,60mm){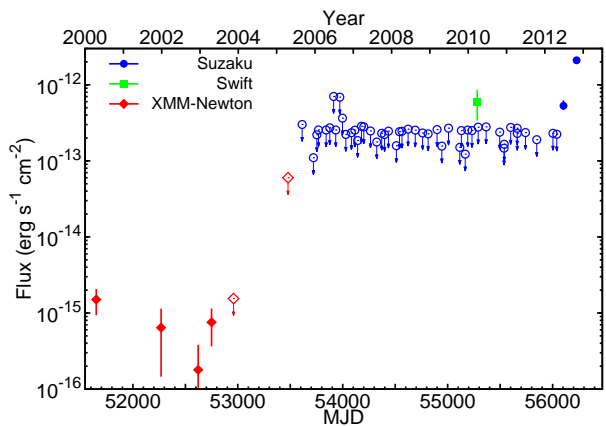}
 \end{center}
 \caption{Long-term change in the 0.5--10~keV flux. Upper limits (5\,$\sigma$) are
 shown by open symbols with downward arrows. The XIS limits were derived in the
 2--10~keV band to avoid readout streaks from the soft and bright source E0102.}
 \label{f6}
\end{figure}

2XMM\,J010247.4--720449 is a B\textit{e}/NS binary candidate from the X-ray spectral
hardness and the association with an early-type star \citep{sturm11}. The occasional
outburst is one of the characteristics of such sources. Indeed, this source was detected
once in outburst by Swift in 2010 March 27, in addition to six times at quiescence
during the SMC survey by XMM-Newton, and also in the stacked Chandra image at quiescence
\citep{sturm11}. During the Swift outburst, the flux was $(6.0 \pm 2.6) \times
10^{-13}$~erg~s$^{-1}$~cm$^{-2}$ in 0.2--10.0~keV, whereas at quiescence, the flux was
2--3$\times$10$^{-14}$~erg~s$^{-1}$~cm$^{-2}$ \citep{sturm11}. The long-term trend
combining all these data (figure~\ref{f6}) shows a drastic flux change of the source by
$\sim$4 orders of magnitude.

During the Swift outburst in 2010, the X-ray spectrum was fitted with an absorbed
power-law model with $\Gamma = 0.9 \pm 0.5$ and $N_{\mathrm{H}}^{\mathrm{SMC}} < 1.5
\times 10^{22}$~cm$^{-2}$. In comparison with our results (table~\ref{t1}), the
power-law index is consistent among the three measurements, while the SMC absorption is
significantly higher in our October result, suggesting that the circumstellar extinction
was higher during the outburst of that epoch.

\citet{sturm11} identified the optical and near-infrared counterpart at the position
consistent with the X-ray source, and their magnitudes are (\textit{U}, \textit{B},
\textit{V}, and \textit{I}) $=$ (14.69, 15.75, 16.00, and 16.30)~mag by
\citet{zaritsky02} and (\textit{J}, \textit{H}, and \textit{K}$_{\mathrm{s}}$) $=$
(16.58, 16.52, and 16.61)~mag by \citet{kato07}, respectively. \citet{sturm11} also
retrieved the \textit{I}-band light curve of the source in the optical gravitational
lensing experiment, in which the magnitude was constant at $\sim$16.25~mag and
brightened to $\sim$15.7~mag on 2009 May 2. \citet{haberl12b} reported detection of an
H$\alpha$ signature in emission with an equivalent width of --4~\AA\ on 2012 December
8. The magnitude, color, brightening trend, and presence of the H$\alpha$ emission line
confirm the hypothesis that this star exhibits a B\textit{e} phenomena.


In summary, the multi-wavelength characteristics are quite typical of those for the
high-mass X-ray binary pulsars with a B\textit{e} star companion in the SMC
\citep{coe05,laycock10}. Our detection of a coherent pulse signal from the source
concludes that this is the case.

\medskip

We acknowledge D. Takei for his help in the XIS astrometric correction. This research
made use of data obtained from Data ARchives and Transmission System (DARTS) by PLAIN
center at ISAS/JAXA, and the SIMBAD database operated at CDS, Strasbourg, France.

\bibliographystyle{aa}
\bibliography{ms}

\end{document}